\documentclass[a4paper,fleqn,usenatbib]{mnras}

\usepackage{newtxtext,newtxmath}
\usepackage[T1]{fontenc}
\usepackage{ae,aecompl}

\usepackage{graphicx}	% Including figure files
\usepackage{amsmath}	% Advanced maths commands
\usepackage{amssymb}	% Extra maths symbols

\title[$\pi^5$ Ori]{BRITE-Constellation photometry of $\bpi^5$ Orionis, an ellipsoidal SPB variable\thanks{Based on data collected by 
the BRITE Constellation satellite mission, designed, built, launched, operated, and supported by the Austrian Research Promotion Agency 
(FFG), the University of Vienna, the Technical University of Graz, the University of Innsbruck, the Canadian Space Agency (CSA), the 
University of Toronto Institute for Aerospace Studies (UTIAS), the Foundation for Polish Science \& Technology (FNiTP MNiSW), and the 
Polish National Science Centre (NCN).}}
\author[M. Jerzykiewicz et al.]{M.~Jerzykiewicz,$^{1}$\thanks{E-mail: jerzykiewicz@astro.uni.wroc.pl} 
A.~Pigulski,$^{1}$ G.~Handler,$^{2}$ A.F.J.~Moffat,$^{3}$ A.~Popowicz,$^{4}$\newauthor G.A.~Wade,$^{5}$ K.~Zwintz$^{6}$ and H.~Pablo$^{7}$
\\  
$^{1}$ Astronomical Institute of the Wroc{\l}aw University, Kopernika 11, 51-622 Wroc{\l}aw, Poland\\ 
$^{2}$ Copernicus Astronomical Center, Bartycka 18, 00-716, Warsaw, Poland\\
$^{3}$ D\'epartement de physique, Universit\'e de Montr\'eal, Canada\\
$^{4}$ Department of Electronics, Electrical Engineering and Microelectronics, Silesian University of Technology, Akademicka 16, 44-100 Gliwice, Poland\\
$^{5}$ Department of Physics and Space Science, Royal Military College of Canada, PO Box 17000, Station Forces, Kingston, ON, Canada, K7K 7B4\\
$^{6}$ Universit\"at Innsbruck, Institute for Astro- and Particle Physics, Technikerstrasse 25, A-6020 Innsbruck\\
$^{7}$ American Association of Variable Star Observers, 49 Bay State Road, Cambridge, MA 02138, USA
}
\date{Accepted XXX. Received YYY; in original form ZZZ}

\pubyear{2020}

\begin{document}
\label{firstpage}
\pagerange{\pageref{firstpage}--\pageref{lastpage}}
\maketitle

\begin{abstract} 
Results of an analysis of the BRITE-Constellation photometry of the SB1 system and ellipsoidal variable $\pi^5$ Ori (B2\,III) are 
presented. In addition to the orbital light-variation, which can be represented as a five-term Fourier cosine series with the 
frequencies $f_{\rm orb}$, $2f_{\rm orb}$, $3f_{\rm orb}$, $4f_{\rm orb}$ and $6f_{\rm orb}$, where $f_{\rm orb}$ is the system's 
orbital frequency, the star shows five low-amplitude but highly-significant sinusoidal variations with frequencies $f_i$ ($i 
={}$2,..,5,7) in the range from 0.16 to 0.92~d$^{-1}$. With an accuracy better than 1$\sigma$, the latter frequencies obey the 
following relations: $f_2-f_4 = 2f_{\rm orb}$, $f_7 - f_3 = 2f_{\rm orb}$, $f_5 = f_3 - f_4 = f_7 - f_2$. We interpret the first 
two relations as evidence that two high-order $\ell = 1, m = 0$ gravity modes are self-excited in the system's tidally distorted 
primary component. The star is thus an ellipsoidal SPB variable. The last relations arise from the existence of the first-order 
differential combination term between the two modes. Fundamental parameters, derived from photometric data in the literature and 
the {\em Hipparcos\/} parallax, indicate that the primary component is close to the terminal stages of its main sequence (MS) 
evolution. Extensive Wilson-Devinney modeling leads to the conclusion that best fits of the theoretical to observed light-curves 
are obtained for the effective temperature and mass consistent with the primary's position in the HR diagram and suggests that the 
secondary is in an early MS evolutionary stage. 
\end{abstract}

\begin{keywords}
stars: early-type --  stars: individual: $\pi^5$ Orionis --  stars: ellipsoidal -- stars: oscillations -- binaries: spectroscopic 
\end{keywords}

\section{Introduction} \label{intro}

The radial velocity (RV) of $\pi^5$ Ori (HD\,31237, HR\,1567, HIP\,22797) was discovered to be variable with a range of about 
110\,km\,s$^{-1}$ by \citet{FrAd03}. \citet*{Lee13} found the star to be a single-lined spectroscopic binary, derived an orbital 
period $P_{\rm orb} ={}$3.70045~d and computed orbital elements assuming zero eccentricity. According to this author ``The lines 
are often faint and always diffuse and difficult to measure. No evidence of the spectrum of the second component has been found.'' 
A single MK type of B2\,III was assigned to the star by \citet*{Lesh68}. However, in {\em The Bright Star Catalogue} \citep{HW91} 
the MK type is given as B3\,III+B0\,V but without any reference. In our opinion, this classification is erroneous: if it were 
correct, the secondary component would be about a magnitude brighter than the primary \citep[see e.g.~table 6 of][]{PCK63}, in 
striking conflict with \citet{Lee13} observation just quoted. \citet{Lee13} elements were refined by \citet*{Mi50} who obtained 
$P_{\rm orb} ={}$3.700373$\,\pm\,$0.000005 d, $K ={}$60.41$\,\pm\,$1.88\,km\,s$^{-1}$, $\gamma 
={}$21.47$\,\pm\,$1.34\,km\,s$^{-1}$, $e ={}$0.073$\,\pm\,$0.040, $\omega ={}$161$\fdg$8$\,\pm\,$47$\fdg$5, $T 
={}$JD\,2433341.088$\,\pm\,$0.019 and $a \sin i ={}$3.07$\times$10$^6$~km. From archival data, \citet{Mo80} derived $e 
={}$0.023$\,\pm\,$0.022 and listed $\pi^5$ Ori among systems with insignificant eccentricity. \citet{Ste20} discovered $\pi^5$ Ori 
to be variable in brightness and classified it as an ellipsoidal variable, the first one of this type ever found. He fitted his 25 
observations with a sine-curve of one-half the orbital period and an amplitude of 0.0267$\,\pm\,$0.0021~mag; the standard 
deviation of the fit amounted to 0.007~mag. The light-variability and the variability classification were confirmed by 
\citet{WR83}. \citet*{Mo85}  solved the ellipsoidal light-curve for two values of the relative brightness of the secondary, a 
primary mass of 8~M$_{\sun}$, and synchronous rotation using \citet{Kop59} Fourier cosine expansion.

\section{The data}\label{data}

The photometry analysed in the present paper was obtained from space by the constellation of BRITE (BRIght Target Explorer) 
nanosatellites \citep{BRITE1,Pab+16} during six observing seasons. The observations were taken in the fields Orion I to V and 
Orion-Taurus I by all five BRITEs, three with red filters: UniBRITE (UBr), BRITE-Toronto (BTr), and BRITE-Heweliusz (BHr), and two 
with blue filters: BRITE-Austria (BAb) and BRITE-Lem (BLb). Details of the observations are given in Table~\ref{Tab-01}. The 
Ori\,I and II observations were obtained in ``stare'' mode, i.e.\ the satellite stabilized mode, the remaining ones, in 
``chopping'' mode, i.e.\ with the satellite moved between two alternative directions to mitigate the problem of defective pixels 
\citep{Pab+16,Pop+17}. The images were analyzed by means of the two pipelines described by \cite{Pop+17}. The resulting aperture 
photometry is subject to several instrumental effects \citep{2018adlc106} and needs post-processing aimed at their removal. In 
order to remove the instrumental effects we followed the procedure of \cite{AP+16} with several modifications proposed by 
\cite{2018adlc175}. The whole procedure includes converting fluxes to magnitudes, rejecting outliers and the worst orbits (i.e.\ 
the orbits on which the standard deviation of the magnitudes, $SD_{\rm sat}$, was excessive), and one- and two-dimensional 
decorrelations with all parameters provided with the data (e.g.~position of the stellar profile in the image or CCD temperature) 
and the calculated satellite orbital phase. Since the Orion fields are rather close to the ecliptic, a number of observations were 
affected by stray light from the Moon; these observations were rejected.

\begin{table*}
\caption{Details of the BRITE data for $\pi^5$~Ori. $N_{\rm orb}$ is the mean number of data points per satellite orbit; $N_{\rm 
orig}$ and $N_{\rm final}$ are the original and final (after post-processing) numbers of data points. RSD is the residual standard 
deviation after subtracting the star's intrinsic variability according to the nine-frequency fits of Section~\ref{FrAn}.}
\label{Tab-01}
\begin{tabular}{cccccrccrrc}
\hline
 Field & Satellite &  Start & End & \multicolumn{1}{c}{Length of}& \multicolumn{1}{c}{Exposure}& \multicolumn{1}{c}{$N_{\rm orb}$} 
& \multicolumn{1}{c}{$N_{\rm orig}$} & \multicolumn{1}{c}{$N_{\rm final}$} & \multicolumn{1}{c}{RSD}& \multicolumn{1}{c}{Nyquist} 
\\ &                  &     date   &  date   & \multicolumn{1}{c}{the run [d]}& \multicolumn{1}{c}{time [s]}  
&&&&\multicolumn{1}{c}{[mmag]}& \multicolumn{1}{c}{frequency [d$^{-1}$]}\\
\hline
 Ori I & BAb & 2013.12.01 & 2014.03.17 & 105.7& 1&36 & 24\,177 & 22\,838  & 10.18&14.35\\
  & UBr & 2013.11.07 & 2014.03.17 & 130.2& 1&45 & 35\,445 & 31\,889  & 12.83&14.35\\
 Ori II  & BAb & 2014.09.25 & 2014.11.08 & 32.7& 1&29 & 5\,671 & 4\,988  & 14.65&14.35\\
  & BLb & 2014.12.07 & 2015.03.16 & 99.6& 1&32 & 34\,010 & 32\,055  & 9.34&14.45\\
  & BTr & 2014.09.24 & 2014.12.04 & 70.8& 1&47 & 31\,836 & 27\,409  & 5.97&14.66\\
  & BHr & 2014.11.10 & 2015.03.14 & 123.3& 1&29 & 34\,293 & 30\,074  & 9.59&14.83\\
Ori III  & UBr & 2015.12.19 & 2016.02.24 & 67.7& 1&25 & 16\,545 & 13\,993  & 13.04&14.35\\
Ori IV  & UBr & 2016.09.13 & 2017.03.01 & 168.8& 1&32 & 28\,337 & 22\,251  & 12.99&14.35 \\
Ori V  & UBr & 2017.09.25 & 2018.02.28 & 155.9& 2&27 & 16\,655 & 11\,366  & 12.29&14.35\\
OriTau I & BAb& 2018.09.13& 2019.03.09& 176.6& 1&17 & 17\,926& 8\,237 & 17.19&14.35\\
  & BLb& 2018.10.08& 2019.03.18& 161.5& 2&27 & 69\,592& 48\,164 & 11.28&14.45\\
 \hline
\end{tabular}
\end{table*}

\section{Frequency analysis}\label{FrAn}

\begin{figure*}   
\includegraphics[width=0.85\textwidth]{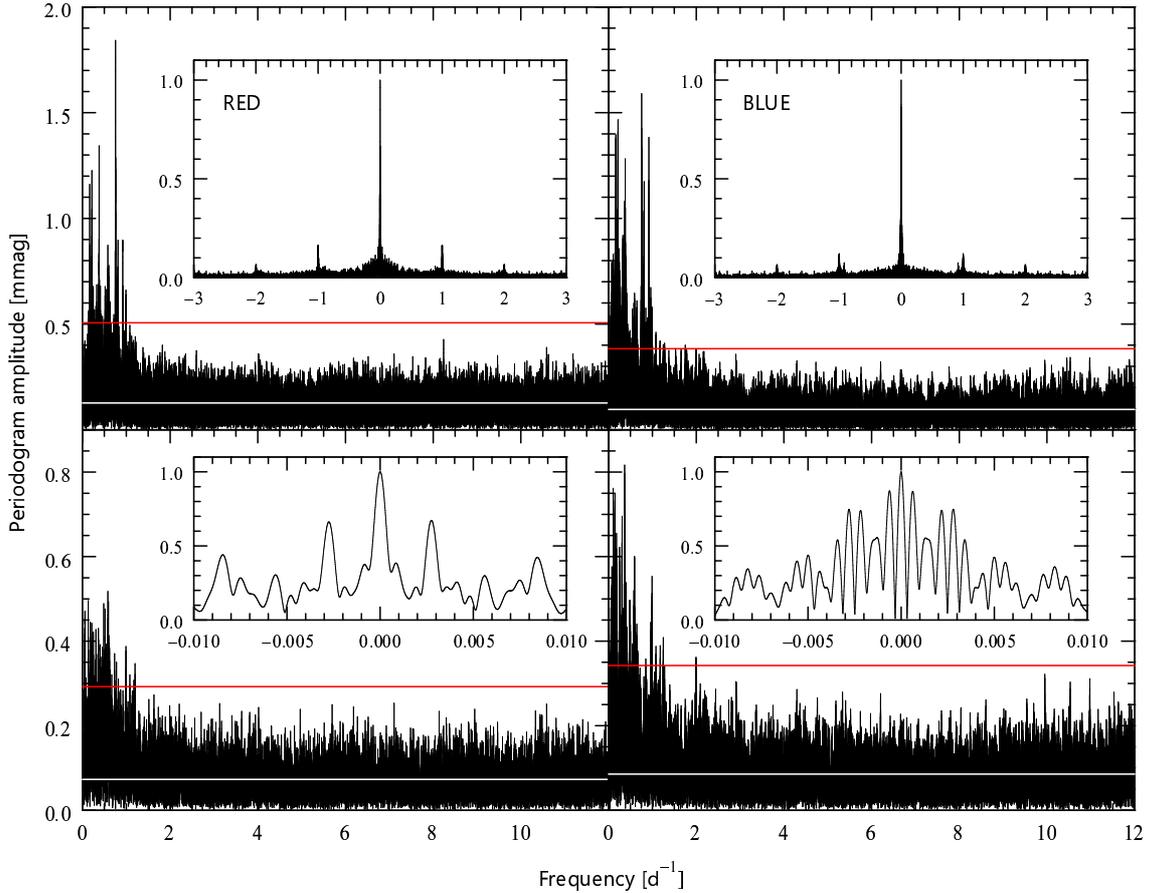} 
\caption{The amplitude spectra of the red and blue BRITE magnitudes pre-whitened with 2$f_{\rm orb}$ and of the residuals from the 
nine-frequency nonlinear least-squares fits (the upper and lower panels, respectively). The mean noise levels, computed from the 
amplitudes in the frequency range from 3.5 to 12~d$^{-1}$, are plotted with the white thick lines, and four times those, with the 
red lines. The spectral windows in the frequency range from $-$3 to 3~d$^{-1}$ are shown in the upper insets, and their central 
lobes, in the lower insets.}
\label{Fig01}    
\end{figure*}

For the purpose of frequency analysis, the reduced UBr, BTr and BHr magnitudes were combined into one set of red magnitudes, and 
the reduced BAb and BLb magnitudes into one set of blue magnitudes. The red magnitudes contained 136\,982 data-points, spanning an 
interval of 1\,574~d; the blue magnitudes contained 116\,282 data-points, spanning 1\,933~d. Thus, the frequency resolution is 
0.0006 and 0.0005~d$^{-1}$ for the red and blue data, respectively. Using these data, we computed the red and blue amplitude 
spectra in the frequency range from 0 to 12~d$^{-1}$. In the process, we applied weights to the magnitudes. The weights were equal 
to $(minSD_{\rm sat}/SD_{\rm sat})^2$, where $minSD_{\rm sat}$ is the smallest value of $SD_{\rm sat}$, the standard deviation of 
the magnitudes in a given orbit. In both cases, the highest peak occurred at 2$f_{\rm orb} ={}$2/$P_{\rm orb}$, where $P_{\rm 
orb}$ is \citet{Mi50} orbital period, to within less than 0.03 of the frequency resolution of the data. The amplitude spectra of 
the red and blue magnitudes pre-whitened with 2$f_{\rm orb}$ are shown in the upper panels of Fig.~\ref{Fig01}. The highest peak 
in both panels is at 0.7560~d$^{-1}$. After pre-whitening the data with 2$f_{\rm orb}$ and the latter frequency, we computed the 
third amplitude spectrum and derived the third frequency of maximum amplitude, etc. The first seven peaks of maximum amplitude 
(including the two mentioned above) in the red amplitude spectra occurred at the same frequencies, or very nearly so, as their 
counterparts in the blue amplitude spectra. In the order of decreasing red amplitude, we shall refer to these frequencies as $f_i$ 
($i ={}$1,..,7). In the eighth, red amplitude spectrum, the two highest peaks of almost the same height appeared at the 
frequencies of 0.27322 and 0.32587~d$^{-1}$. The highest peak in the blue spectrum occurred at the latter frequency; we shall 
refer to this frequency as $f_8$. The former frequency is close to that of a sidereal year alias of $f_{\rm orb}$; the alias is 
present in the red and blue spectra at 0.2703~d$^{-1}$. We shall refer to this frequency as $f_9$. In order to refine the nine 
frequencies, we fitted the red and blue magnitudes with the equation 
\begin{equation} 
{\rm mag} = A_0 + \sum_{i=1}^{9} A_i \cos [2 \pi f_i ({\rm HJD} - 2456900) + \phi_i], 
\end{equation} 
by means of the method of nonlinear least-squares \citep{Sch1908} using the frequencies derived from the amplitude spectra as 
starting values and the same weights as in computing the amplitude spectra. Results are presented in Table~\ref{Tab-02}. The $SD$ 
in the heading are the standard deviations of the right-hand side of the observational equation of unit weight. The frequencies, 
$f_i$ ($i ={}$1,..,9), listed in column two are weighted means of those from the red and blue solutions. In the frequency analysis 
of extensive photometric time-series of XX Pyx \citep{H+00} and that of $\nu$ Eri \citep{jhs}, the formal least-squares standard 
deviations of $f_i$, $A_i$ and $\phi_i$ were found to be underestimated by a factor of about two. We believe that this also 
applies to the standard deviations in Table~\ref{Tab-02}. Columns five and eight of the table contain the signal-to-noise ratio, 
$S/N$, where $S$ is the amplitude and $N$ is the mean level of noise estimated as explained in the caption to Fig.~\ref{Fig01}. In 
all cases $S/N >{}$4, the popular detection threshold set by \citet{B+93}.

\begin{table*}
\caption{The parameters of a nonlinear least-squares fit of equation (1) to the red and blue BRITE magnitudes.}
\label{Tab-02}
\begin{center}
RED:\ $n ={}$136\,982, $SD ={}$0.29 mmag, $A_0 ={}$0.01$\,\pm\,$0.02 mmag. \ BLUE:\ $n ={}$116\,282, $SD ={}$1.41 mmag, $A_0 
={}$0.00$\,\pm\,$0.03 mmag
  \begin{tabular}{@{}lcrcrrcr@{}}
  \hline
& & \multicolumn{3}{c}{RED}  &  \multicolumn{3}{c}{BLUE}  \\
 \multicolumn{1}{c}{$i$}&\multicolumn{1}{c}{$f_i$ [d$^{-1}$]}&\multicolumn{1}{c}{$A_i$ [mmag]} & \multicolumn{1}{c}{$\phi_i$ 
[rad]}&S/N&\multicolumn{1}{c}{$A_i$ [mmag]} & \multicolumn{1}{c}{$\phi_i$ [rad]}&S/N\\
  \hline
   1=$2f_{\rm orb}$&\hspace{4pt}0.5404851$\pm$0.0000014&23.13$\pm$0.03&4.6147$\pm$0.0016 &316.4&24.13$\pm$0.04&4.6185$\pm$0.0022&282.2\\
   2&\hspace{4pt}0.7559594$\pm$0.0000062&1.95$\pm$0.03&0.117$\pm$0.019&26.7& 1.63$\pm$0.04&0.149$\pm$0.033&19.1\\
   3&\hspace{4pt}0.379780$\pm$0.000041&1.43$\pm$0.03&4.751$\pm$0.026&19.6& 1.14$\pm$0.04&4.857$\pm$0.046&13.3\\
   4&\hspace{4pt}0.215476$\pm$0.000025&1.30$\pm$0.03&1.779$\pm$0.028&17.8& 1.59$\pm$0.04&1.598$\pm$0.034&18.6\\
   5&\hspace{4pt}0.164293$\pm$0.000015&1.26$\pm$0.03&2.959$\pm$0.030&17.2& 1.33$\pm$0.04&2.715$\pm$0.040&15.6\\
   6=$3f_{\rm orb}$&\hspace{4pt}0.810723$\pm$0.000010&1.04$\pm$0.03&0.702$\pm$0.036&14.2& 1.40$\pm$0.04&0.533$\pm$0.038&16.4\\
   7&\hspace{4pt}0.920221$\pm$0.000012&1.03$\pm$0.03&3.152$\pm$0.036&14.1& 1.40$\pm$0.04&3.056$\pm$0.037&16.4\\
   8&\hspace{4pt}0.325873$\pm$0.000019&0.70$\pm$0.03&1.799$\pm$0.052&9.6& 0.92$\pm$0.04&2.220$\pm$0.056&10.8\\
   9=$f_{\rm orb}$&\hspace{4pt}0.27030$\pm$0.00012&0.63$\pm$0.03&2.181$\pm$0.058&8.6&0.40$\pm$0.04&4.85$\pm$0.13&4.7\\
  \hline
\end{tabular}
\end{center}
\end{table*}

The amplitude spectra of the red and blue residuals from the nine-frequency nonlinear least-squares fits are plotted in the lower 
panels of Fig.~\ref{Fig01}. The numerous peaks higher than 4$N$ and a gradual increase of the mean level of the signal at 
frequencies lower than about 3~d$^{-1}$ with decreasing frequency seen in the amplitude spectra of the residuals from the 
9-frequency fits (lower panels of the figure) are peculiar to $\pi^5$ Ori. The amplitude spectra of the BRITE magnitudes of 
several other stars observed under similar circumstances and reduced in the same way as $\pi^5$ Ori are flat throughout. An 
example is the B0.5\,IV eclipsing variable $\delta$ Pic. The amplitude spectrum of the BHr magnitudes of $\delta$ Pic with the 
eclipsing light-variation removed, seen in fig.~2 of \citet{2017adlc120}, shows no amplitude increase with decreasing frequency. 
Two further examples are HR\,6628 and $\pi$ Cen. HR\,6628, a 4.8 mag B8\,V star, was observed in 2017 and 2018. Apart from a 
single $S/N ={}$4.2 peak at the frequency of 0.0445~d$^{-1}$, the 0 to 12~d$^{-1}$ amplitude spectrum of the combined 85096 BLb 
and 11888 BAb magnitudes is flat, with the mean level of noise $N ={}$0.16~mmag. Frequency analysis of the combined 9239 BLb and 
65235 BTr 2016 magnitudes of $\pi$ Cen (3.9~mag, B5\,Vn) yielded six sinusoidal terms with frequencies in the range 2.27 to 
5.21~d$^{-1}$ with amplitudes 0.72 to 1.84~mmag. The 0 to 12~d$^{-1}$ amplitude spectrum after pre-whitening with these terms was 
flat, with no peaks higher than 0.30~mmag and $N ={}$0.08~mmag. Returning to $\pi^5$ Ori, we conclude from the behaviour of the 
amplitude spectra at low frequencies that in addition to the two $\ell = 1, m = 0$ gravity modes identified in Section~\ref{disc}, 
other low-frequency, $\ell \geq{}$1 gravity modes are excited in the primary component of $\pi^5$ Ori. As discussed in 
Section~\ref{disc}, each $\ell, m$ frequency would be split in the observer's frame into several frequencies. The amplitude 
spectra in Fig.~\ref{Fig01} are the result of an interference of the spectral windows shifted to the positions of the frequencies 
and scaled by the corresponding amplitudes. In addition, negative-frequency signals leaking to the positive-frequency domain 
contribute to the interference. Unfortunately, the spectral windows are rather complex and do not match each other. As can be seen 
from the lower insets in Fig.~\ref{Fig01}, the single central peak of the red-band spectral window is replaced in the blue-band 
spectral window by three peaks of almost the same amplitude. It is thus not surprising that the red and blue amplitude spectra of 
the residuals do not match. An attempt to reveal an $i >{}$9 frequency common to the red and blue frequency spectra of the 
residuals was unsuccessful. We therefore decided to terminate the frequency analysis at this stage. 

\begin{figure}   
\includegraphics[width=\columnwidth]{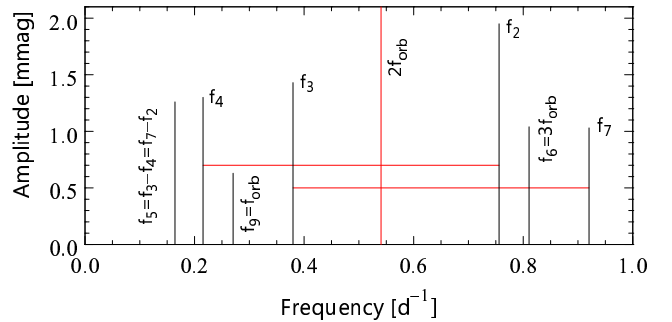} 
\caption{Schematic frequency spectrum of $\pi^5$ Ori plotted using the frequencies and the red amplitudes from Table~\ref{Tab-02}. 
The $2f_{\rm orb}$ amplitude is off scale. The horizontal red lines represent the $2f_{\rm orb}$ separation between the doublets 
$f_4,f_2$ and $f_3,f_7$.}
\label{Fig02}    
\end{figure}

As can be seen from Table~\ref{Tab-02}, the frequencies $f_i$ ($i ={}$2,..,7,9) are related to each other and to $2f_{\rm orb}$: 
\begin{eqnarray*}
f_2-f_4 - 2f_{\rm orb} = -0.000030\pm0.000026\ {\rm d}^{-1}, \\
f_7-f_3 - 2f_{\rm orb} = -0.000045\pm0.000043\ {\rm d}^{-1}, \\
f_3-f_4 -f_5 ={} 0.000010\pm0.000051\ {\rm d}^{-1}, \\
f_6-3 f_{\rm orb} = -0.000006\pm0.000010\ {\rm d}^{-1},\\
f_9-f_{\rm orb} = 0.00006\pm0.00012\ {\rm d}^{-1},\\
\end{eqnarray*}
where the standard deviations were computed from the underestimated formal standard deviations of Table~\ref{Tab-02}. Thus, with 
an accuracy better than 1$\sigma$ these interconnections lead to the following relations:
\begin{eqnarray}
f_2 - f_4 = 2f_{\rm orb}, \\
f_7 - f_3 = 2f_{\rm orb}, \\
f_5 = f_3 - f_4, \\
f_6 = 3f_{\rm orb},\\
f_9 = f_{\rm orb},
\end{eqnarray}
illustrated in Fig.~\ref{Fig02}. Note that equation (4) can be replaced by 
\begin{equation}
f_5 = f_7 - f_2,
\end{equation}
while equations (2) and (3) lead to
\begin{equation}
f_2 + f_3 = f_4 + f_7.
\end{equation}

\section{The orbital light and RV curves}\label{orbital-curves}

\begin{figure} 
\includegraphics[width=\columnwidth]{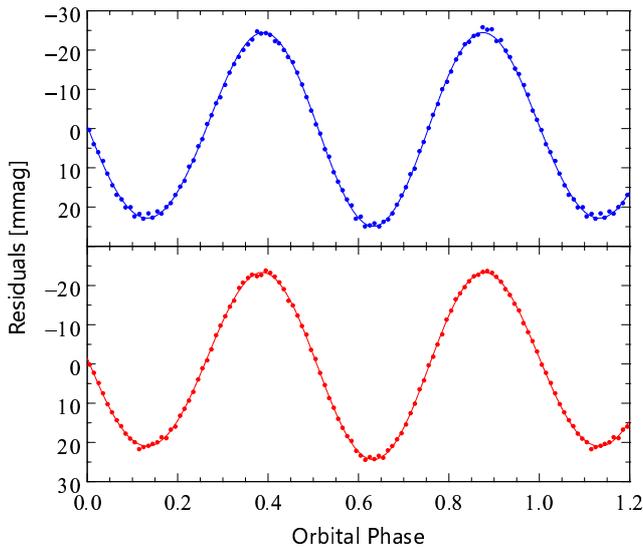} 
\caption{Normal points (dots), computed from the red (lower panel) and blue (upper panel) magnitudes pre-whitened with the $f_i$ 
($i ={}$2,..,5,7,8) terms using the parameters of the red and blue nonlinear least-squares fits of Section~\ref{FrAn}. The epoch 
of phase zero is HJD\,2456900. The lines are the theoretical light-curves computed from the W-D solutions detailed in the text. 
The theoretical light-curves fit the normal points with standard deviations of 0.40 and 0.56~mmag for the red and blue data, 
respectively.}
\label{Fig03} 
\end{figure}

\begin{figure} 
\includegraphics[width=\columnwidth]{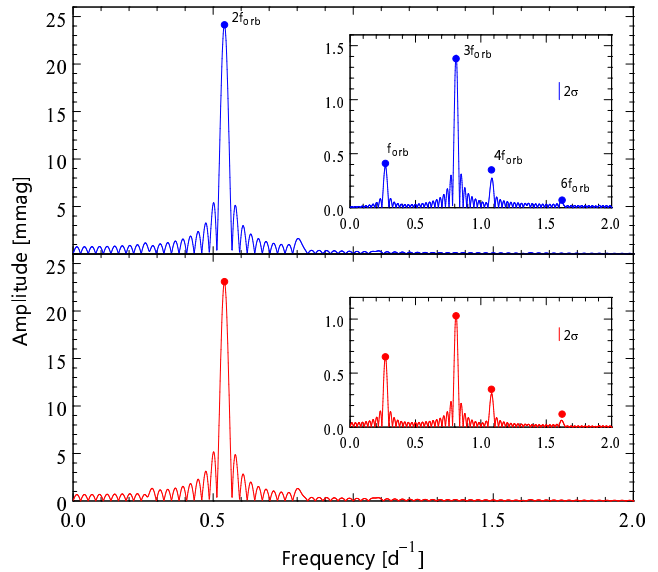} 
\caption{A frequency-domain counterpart of Fig.~\ref{Fig03}. The circles are the amplitudes of the five-term Fourier-series 
least-squares fits to the normal points. The error bars are labeled 2$\sigma$, where $\sigma$ is the formal least-squares standard 
deviation of the amplitudes. The lines in the large panels are the amplitude spectra computed from the W-D theoretical 
light-curves of ten cycles, while those in the insets, from the theoretical light-curves pre-whitened with the 2$f_{\rm orb}$ 
term.}
\label{Fig04} 
\end{figure}

The red- and blue-magnitude phase-diagrams are plotted in Fig.~\ref{Fig03}. The phases were computed with \citet{Mi50} orbital 
period of 3.700373~d and the epoch of phase zero HJD\,2456900. The data shown as dots are normal points, formed in adjacent 
intervals of 0.01 orbital phase from the red and blue magnitudes pre-whitened with the $f_i$ ($i ={}$2,..,5,7,8) terms using the 
parameters of the red and blue nonlinear least-squares fits of Section~\ref{FrAn}. Error bars are not plotted because they would 
rarely extend beyond the dots: the standard errors ranged from 0.10 to 0.28~mmag for the red normal points, and from 0.23 to 
0.31~mmag for the blue normal points. The lines are the theoretical light-curves, computed from a Wilson-Devinney (W-D) solution 
obtained under assumption of synchronous rotation using the observed $V_{\rm rot} \sin i ={}$90~km\,s$^{-1}$ \citep{GG05} and 
assuming the parameters $R_1 ={}$11.6~R$_{\sun}$, $M_1 ={}$12.0~M$_{\sun}$, $T_{\rm eff,1} ={}$21\,590~K for the primary 
component, and  $R_2 ={}$2.83~R$_{\sun}$, $M_2 ={}$4.95~M$_{\sun}$, $T_{\rm eff,2} ={}$16\,500~K, and the radiative-envelope 
bolometric albedo $\alpha_2 ={}$1.0 for the secondary component, i.e.\ the first solution in Table~\ref{TabB3}. The W-D phase of 
the deeper minima is 0.6325. The depth difference between minima is equal to 3.5 and 2.0~mmag for the red and blue light-curves, 
respectively. In the W-D solutions, the reflection effect accounts for 2.1~mmag of the red depth difference and the entire blue 
depth difference. The W-D modeling will be discussed in Section~\ref{WDsolutions}. Figure~\ref{Fig04} is a frequency-domain 
counterpart of Fig.~\ref{Fig03}. In Fig.~\ref{Fig04}, the lines in the large panels are the amplitude spectra computed from the 
theoretical light-curves of ten cycles, while those in the insets, from the theoretical light-curves pre-whitened with the 
2$f_{\rm orb}$ term. The circles are the amplitudes of the five-term Fourier-series least-squares fits to the normal points. The 
$f_4$ and $f_6$ terms were included in the fit so that their amplitudes could be compared with the theoretical ones. In both 
bands, the observed and theoretical amplitudes agree very well with each other. 

\begin{figure} 
\includegraphics[width=\columnwidth]{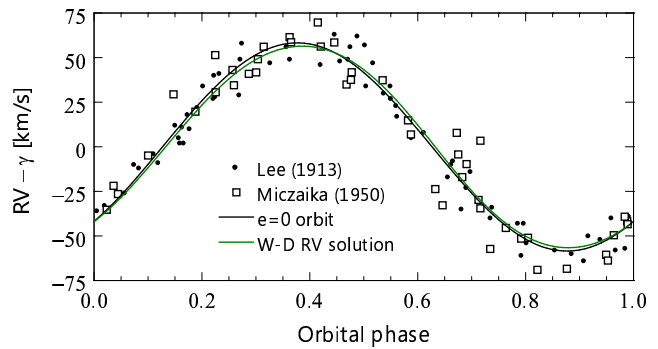} 
\caption{The archival RVs of $\pi^5$ Ori plotted as a function of phase of the orbital period. The epoch of phase zero is HJD\,2456900. 
An $e ={}$0 orbital RV curve and the RV curve from the W-D solution mentioned in the text are shown as the black and green line, 
respectively.}
\label{Fig05} 
\end{figure}

Archival RVs of $\pi^5$ Ori are plotted in Fig.~\ref{Fig05} together with an $e ={}$0 orbital RV curve and the RV curve from the 
W-D solution mentioned in the preceding paragraph. The amplitude of the $e ={}$0 curve $K_1 ={}$58.4$\pm$1.3~km/s and the phase of 
the minimum is equal to 0.879$\pm$0.004. The difference between the latter number and the above-mentioned phase of the deeper 
minima of the light-curves differs from the expected 0.25 by less than 1$\sigma$. 

\section{Discussion and conclusions}\label{disc}

The system $\pi^5$ Ori is a simple one: the orbit is circular and the components can be safely assumed to rotate synchronously 
\citep[see][and references therein]{Lev76}. Under such circumstances the tidal force does not change, resulting in the so-called 
equilibrium tide in which tidal distortion remains constant and the light-variation is caused by the variation of the projected 
area of the components as a function of phase of $2f_{\rm orb}$. The difference in the depth of the alternate minima seen in 
Fig.~\ref{Fig03} reveals that in the case of $\pi^5$ Ori this ellipsoidal variation is modified by a small but significant 
reflection effect. Under the assumption of synchronous rotation, the best fits of the W-D to the observed light-curves are 
obtained for $M_1 ={}$12~M$_{\sun}$ with $\log T_{\rm eff,1}$ and $\log L_1/$L$_{\sun}$ within 1$\sigma$ of the HR diagram 
position of the star derived in Section~\ref{parameters} from photometric data from the literature and the Hipparcos parallax and 
in limited ranges of $\log T_{\rm eff,2}$, different for the two cases we consider, viz.\ that of a radiative-envelope bolometric 
albedo $\alpha_2 ={}$1.0 and a convective-envelope bolometric albedo $\alpha_2 ={}$0.5 (see Table~\ref{TabB1}). The primary 
component of $\pi^5$ Ori is thus found to be in a more advanced stage of evolution than components of the SB2 eclipsing binaries 
of comparable masses in table 1 of \citet*{TAG10}. Although the magnitude difference between the components is not known, we 
derive duplicity corrections for the two cases of the bolometric albedo of the secondary using magnitude differences from the W-D 
modeling and assuming that $M_2$, the secondary component's mass from the orbital solution is equal to its evolutionary mass (see 
Tables~\ref{TabB2} and \ref{TabB3} and Fig.~\ref{FigA1}). A comparison of the evolutionary age of the secondary with that of the 
primary shows that in the $\alpha_2 ={}$0.5 case the evolutionary age is over an order of magnitude too small, while in the 
$\alpha_2 ={}$1.0 case the difference of the evolutionary masses is probably within the uncertainties of the analysis, suggesting 
that the secondary is in an early stage of its MS evolution (open triangle at lower right in Figs.~\ref{FigA1} and \ref{FigA2}). 

In addition to causing the ellipsoidal light-variation, the equilibrium tide modifies the frequencies of the self-driven 
pulsations of the components. According to the theoretical work of \cite{ReySmey03} and \cite{Smey05}, summarized recently by 
\cite{Bal18}, a non-radial pulsation mode perturbed by an equilibrium tide can be described by a set of independent modes that are 
associated with a single spherical harmonic $Y_\ell^m(\theta,\phi)$ where $\theta$ and $\phi$ are the polar and azimuthal angles 
in a spherical coordinate system whose polar axis coincides with the line joining the components' mass centres. In the corotating 
frame, each $2(\ell+1)$-fold degenerate eigenfrequency of a mode $n$ is split into $\ell+1$ eigenfrequencies. In the non-rotating 
frame with the polar axis parallel to the pulsating component's rotation axis, an $\ell = 1$, $m = 0$ eigenfrequency is split into 
two frequencies, while that of the $\ell = 1$, $m = 1$ eigenfrequency, into three frequencies. To first-order in a small 
dimensionless parameter $\epsilon_T = (R/a)^3 q$, where $R$ is the radius of the pulsating component and $q$ is the mass ratio, 
the two $\ell = 1$, $m = 0$ frequencies, $f^{(1,0)}_1$ and  $f^{(1,0)}_2$ are given by:
\begin{eqnarray}
f^{(1,0)}_1 = f_{n,0} + \epsilon_T f^{(1,0)}_{n,1} - \Omega/2 \pi, \\
f^{(1,0)}_2 = f_{n,0} + \epsilon_T f^{(1,0)}_{n,1} + \Omega/2 \pi,
\end{eqnarray}
while the three $\ell = 1$, $m = 1$ frequencies, $f^{(1,1)}_1$, $f^{(1,1)}_2$ and  $f^{(1,1)}_3$, by:
\begin{eqnarray}
f^{(1,1)}_1 = f_{n,0} + \epsilon_T f^{(1,1)}_{n,1} - \Omega/2 \pi, \\
f^{(1,1)}_2 = f_{n,0} + \epsilon_T f^{(1,1)}_{n,1},\\
f^{(1,1)}_3 = f_{n,0} + \epsilon_T f^{(1,1)}_{n,1} + \Omega/2 \pi,
\end{eqnarray}
where $f_{n,0}$ is the eigenfrequency of the unperturbed mode, $\epsilon_T f^{(\ell,m)}_{n,1}$ are the first order corrections to 
$f_{n,0}$, and $\Omega$ is the angular velocity of rotation. In the case of $\ell = 2$, $|m| \leq \ell$, the $\ell + 1$ 
eigenfrequencies would be split into 12 frequencies that include an equidistant triplet, quadruplet and quintuplet; in the case of 
$\ell = 3$, the eigenfrequencies split into 24 frequencies that include an equidistant quadruplet, sextuplet and two septuplets 
\citep[see table 1 and fig.~2 of][]{Bal18}. In the frequency spectrum of a pulsating component, the frequencies $f^{(1,0)}_1$ and 
$f^{(1,0)}_2$ would form a doublet with separation equal to $\Omega/\pi = 2f_{\rm orb}$, while frequencies $f^{(1,1)}_1$, 
$f^{(1,1)}_2$ and $f^{(1,1)}_3$, an equidistant triplet with the separation equal to $\Omega/2 \pi = f_{\rm orb}$. As can be seen 
from Fig.~\ref{Fig02}, in the frequency spectrum of $\pi^5$ Ori there are two doublets separated by $2f_{\rm orb}$, viz.\ $f_4, 
f_2$ and $f_3,f_7$, but no equidistant triplets. From equations (9)-(13) we conclude that two $l=1, m=0$ modes, $n$ and $n'$, are 
excited in the primary component of $\pi^5$ Ori. Using $R_1$, $M_1$ and $M_2$ from Table~\ref{TabB3}, we get $\epsilon_T < 0.05$. 
Neglecting the second term on the rhs of equations (9) and (10), we obtain approximate values of the unperturbed frequencies, 
$f_{n,0} \approx f_2-f_{\rm orb} = f_4+f_{\rm orb} = (f_2 + f_4)/2 = 0.49$~d$^{-1}$ and $f_{n',0} \approx f_3+f_{\rm orb} = 
f_7-f_{\rm orb} = (f_3 + f_7)/2 = 0.65$~d$^{-1}$. These values of $f_{n,0}$ and $f_{n',0}$ are characteristic of high-order $\ell 
= 1$ gravity modes, so that $\pi^5$ Ori should be classified as an ellipsoidal SPB variable or ELL/LPB(LBV) in the parlance of the 
General Catalogue of Variable Stars\footnote{http://www.sai.msu.su/gcvs/gcvs/}.

The first order combination terms between the modes $n$ and $n'$ have the following frequencies
\begin{eqnarray}
f_{n,0} + f_{n',0} = f_2 + f_3  - \epsilon_T \left(f^{(1,0)}_{n,1} + f^{(1,0)}_{n',1} \right) \nonumber \\ = f_4 + f_7  - 
\epsilon_T \left(f^{(1,0)}_{n,1} + f^{(1,0)}_{n',1} \right)
\end{eqnarray}
and
\begin{eqnarray}
f_{n',0} - f_{n,0} = f_3 - f_4  + \epsilon_T \left(f^{(1,0)}_{n,1} - f^{(1,0)}_{n',1} \right) \nonumber \\ = f_7 - f_2  + 
\epsilon_T \left(f^{(1,0)}_{n,1} - f^{(1,0)}_{n',1} \right).
\end{eqnarray}
Given negligible first-order corrections $\epsilon_T f^{(1,0)}$, equations (15) are consistent with equations (4) and (7), while 
equations (14), with equations (8).

\begin{figure*}
\includegraphics[width=0.85\textwidth]{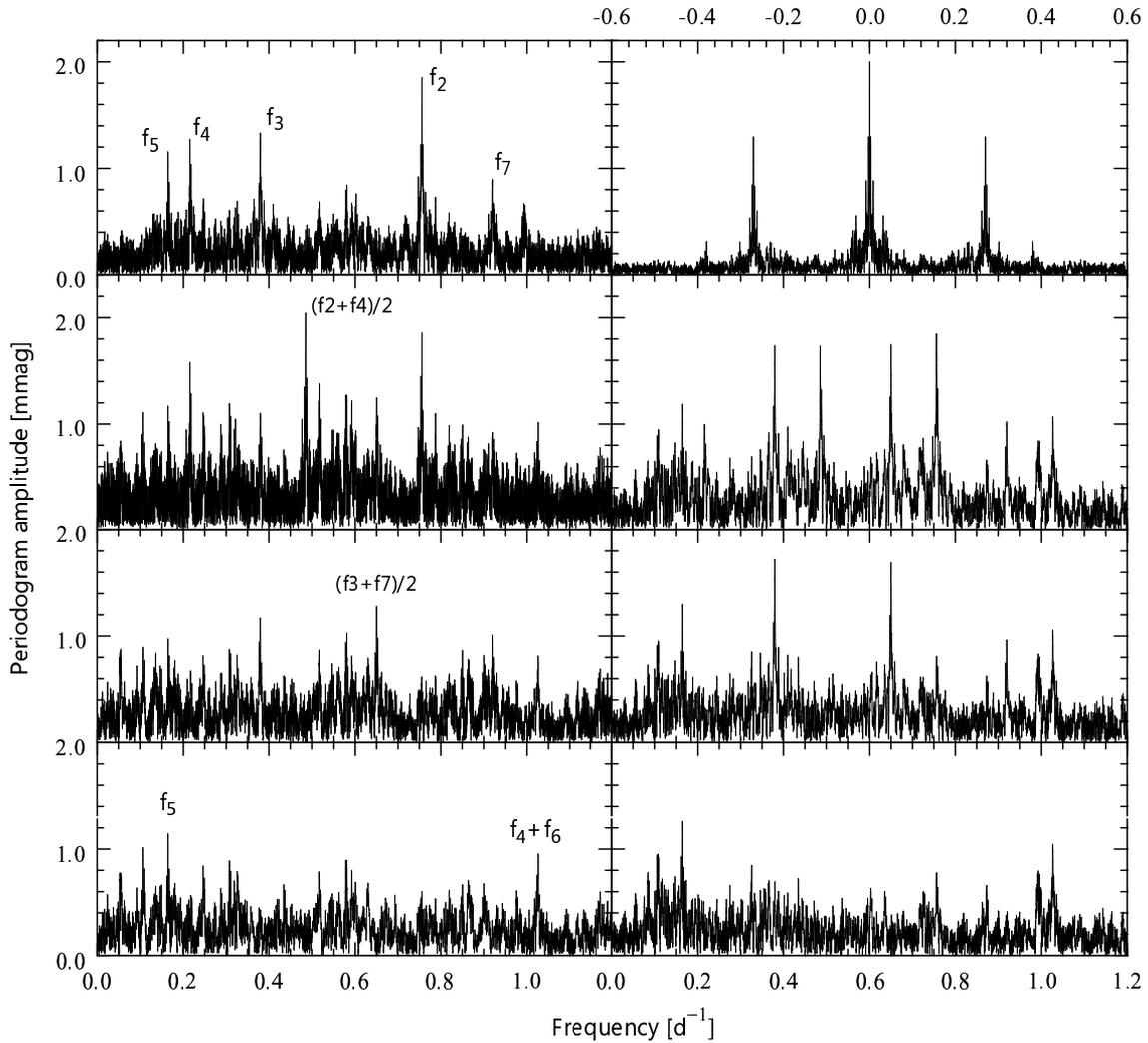}
\caption{The amplitude spectra of three sets of the BRITE red magnitudes of $\pi^5$ Ori pre-whitened with the orbital 
light-variation: (1) all magnitudes (the top left-hand panel), (2) the magnitudes covering orbital phases from the western to 
eastern quadrature, i.e.\ the phases from 0.3825 to 0.8825 in Figs.~\ref{Fig03} and \ref{Fig05} (the remaining left-hand panels), 
and (3) the magnitudes covering the remaining orbital phases, i.e.\ the phases from 0 to 0.3825 and from 0.8825 to 1 in 
Figs.~\ref{Fig03} and \ref{Fig05} (the right-hand panels, from upper middle to bottom). The top right-hand panel shows the 
spectral window of set 3; at the resolution of the figure, the spectral window of set 2 would be very nearly identical with the 
one shown. The lower middle and bottom panels show the amplitude spectra of the set 2 and 3 data pre-whitened with $f_{n,0} = (f_2 
+ f_4)/2 ={}$0.4857~d$^{-1}$, and with this frequency and $f_{n',0} = (f_3 + f_7)/2 ={}$0.6500~d$^{-1}$, respectively.}
\label{Fig06}
\end{figure*}

The referee has suggested a test that our $f_{n,0}$ and $f_{n',0}$ modes are indeed associated with the $\ell = 1$, $m = 
0$ spherical harmonics and provided examples of applying the test to simulated data. The test consists in dividing the data into 
two parts according to the orbital phase in such a way that one part contains the data covering orbital phases from one quadrature 
to the other, and the second part, the remaining data, and then computing amplitude spectra for the two parts separately. Using 
simulated $\ell = 1$, $m = 0$ light-curves with an assumed pulsation frequency, \citet*{RBK05} found for a range of inclination of 
the pulsation axis to the line of sight that in the amplitude spectra of the two parts of the data there appears a peak at the 
assumed frequency flanked by $f_{\rm orb}$ aliases whereas in the amplitude spectrum of the complete data set the peak at the 
assumed frequency is missing (see their figure 4). In addition, there is a phase difference equal to $\pi$ between the 
light-curves in the two parts of the data. For the test, we used the red BRITE magnitudes because their spectral window is cleaner 
than that of the blue magnitudes (see the insets in the lower panels of Fig.~\ref{Fig01}). We removed the orbital light-variation 
by pre-whitening with $f_{\rm orb}$, 2$f_{\rm orb}$, 3$f_{\rm orb}$, 4$f_{\rm orb}$ and  6$f_{\rm orb}$, divided the data into two 
parts as described above, and computed amplitude spectra. The results are displayed in Fig.~\ref{Fig06}. The top left-hand panel 
shows the amplitude spectrum of the complete data (referred to as set 1 in the caption to the figure) with the peaks at the 
frequencies appearing in equations (2)-(5) labelled. The upper middle left-hand panel shows the amplitude spectrum of the 
magnitudes covering the orbital phases from the western to eastern quadrature, i.e.\ the phases from 0.3825 to 0.8825 in 
Figs.~\ref{Fig03} and \ref{Fig05} (set 2). The peak at $f_{n,0} = (f_2 + f_4)/2 ={}$0.4857~d$^{-1}$ and its $f_{\rm orb}$ aliases 
dominate the spectrum. The aliases occur at the same frequencies as the $f_2$ and $f_4$ peaks in the top left-hand panel but should 
not be confused with them. While the aliases reproduce the side-lobes of the spectral window seen in the top right-hand panel, the 
frequencies $f_2$ and $f_4$ arise as the result of a transformation of the corotating frame of reference whose polar axis 
coincides with the line joining the components' mass centres to the non-rotating frame with the polar axis parallel to the 
pulsating component's rotation axis (see the second paragraph of this section). The lower middle left-hand panel contains the 
amplitude spectrum obtained from the set 2 data pre-whitened with $f_{n,0}$. Now, the highest peak appears at $f_{n',0} = (f_3 + 
f_7)/2 ={}$0.6500~d$^{-1}$. Finally, the amplitude spectrum of the set 2 data pre-whitened with $f_{n,0}$ and $f_{n',0}$ is shown 
in the bottom left-hand panel. Here, the two highest peaks appear at 0.1643~d$^{-1}{}= f_5$ and 1.0262~d$^{-1}{}= f_4 + f_6$. The 
amplitude spectra of the second part of the data, i.e.\ the data covering orbital phases from 0 to 0.3825 and from 0.8825 to 1 in 
Figs.~\ref{Fig03} and \ref{Fig05} (set 3) are shown in three right-hand panels. The amplitude spectra in the right-hand middle 
panels differ in appearance from their left-hand counterparts but still the peaks at the frequencies $f_{n,0}$, $f_{n',0}$ and 
their $f_{\rm orb}$ aliases are the strongest features present. The phases of the $f_{n,0}$ and $f_{n',0}$ light-curves computed 
for set 2 and 3 separately are equal to 5.500$\pm$0.022 and 2.689$\pm$0.025~rad for $f_{n,0}$ and 2.524$\pm$0.029 and 
5.351$\pm$0.026~rad for $f_{n',0}$. The phase differences between the light-curves of set 2 and 3 amount to (0.895$\pm$0.011)$\pi$ 
and ($-$0.900$\pm$0.012)$\pi$ for $f_{n,0}$ and $f_{n',0}$, respectively. The outcome of the test is thus mixed: the amplitudes of 
the $f_{n,0}$ and $f_{n',0}$ modes behave as predicted by the $\ell = 1$, $m = 0$ simulations of \citet{RBK05} but the phase 
differences, although close to, are significantly smaller than $\pi$, even if the formal standard deviations were to be 
underestimated by a factor of two as maintained in Section~\ref{FrAn}. 

The highest peak in the bottom left-hand panel of Fig.~\ref{Fig06} at the combination frequency $f_5 = f_{n',0} - f_{n,0}$, 
mentioned earlier in this section, has very nearly the same amplitude in the bottom right-hand and top left-hand panels. One would 
therefore expect that the $f_5$ light-curves of set 2 and 3 will be in phase. In fact, the phases are equal to 4.860$\pm$0.037 and 
4.559$\pm$0.035~rad, so that the light-curves differ in phase by (0.096$\pm$0.016)$\pi$. If we were to take this result as an 
indication that the standard deviations of the phase differences are underestimated by a factor of about six instead of two, the 
deviations of the phase differences from $\pi$ noted at the end of the preceding section would become tolerable. The second 
highest peak in the bottom panels of Fig.~\ref{Fig06} occurs at the frequency $f_4+f_6$. It has no counterpart in the top 
left-hand panel or in the left-hand panels of Fig.~\ref{Fig01}. Now the phase difference between sets 2 and 3 amounts to 
(0.962$\pm$0.020)$\pi$, as one would expect. 

In closing, we would like to venture a prediction: frequencies resulting from the tidal splitting of the the $\ell = 1$, $m = 1$ 
and $\ell = 2$, $|m| \leq \ell$ eigenfrequencies will be eventually identified at the low end of the frequency axis where the 
present analysis failed (see Fig.~\ref{Fig01}).

\section*{Acknowledgments}

In this research, we have used the Aladin service, operated at CDS, Strasbourg, France, and the SAO/NASA Astrophysics Data System 
Abstract Service. A.~Pigulski acknowledges support from the National Science Centre (NCN) grant 2016/21/B/ST9/01126. GH 
acknowledges support by the Polish NCN grant UMO-2015/18/A/ST9/00578. AFJM is grateful for financial aid from NSERC (Canada). 
A.~Popowicz was responsible for image processing and automation of photometric routines for the data registered by 
BRITE-nanosatellite constellation, and was supported by Silesian University of Technology Rector Grant 02/140/RGJ20/0001. GAW 
acknowledges Discovery Grant support from the Natural Sciences and Engineering Research Council (NSERC) of Canada. KZ acknowledges 
support by the Austrian Space Application Programme (ASAP) of the Austrian Research Promotion Agency (FFG). We are indebted to Dr 
M.D.\ Reed, the referee, for suggesting the test described in Section \ref{disc}.

\appendix
\section{Fundamental Parameters}\label{parameters}

Let us start with deriving the colour excess of $\pi^5$ Ori. From the Str\"omgren indices $b-y$ and $c_1$ \citep{hm} we get $c_0 
={}$0.125, $(b-y)_0 ={}-$0.105, $E(b-y) ={}$0.044 and $E(B-V) ={}$0.059~mag by means of the canonical method of \citet*{c78}. From the 
$UBV$ colour indices \citep{m91} and the standard two-colour relation for luminosity class III B-type stars \citep{j63} we get $E(B-V) 
={}$0.058~mag. The excellent agreement of these values of $E(B-V)$ may be somewhat accidental.

We shall now use $c_0$ to estimate the effective temperature, $T_{\rm eff,1}$, and the bolometric correction, $BC_1$, of the primary 
component of $\pi^5$ Ori assuming negligible brightness of the secondary. We get $T_{\rm eff,1} ={}$21\,125~K and $BC_1 ={}-$2.14~mag 
using the calibration of \citet{ds77}, $T_{\rm eff,1} ={}$21\,314~K using {\sc UVBYBETA}\footnote{A FORTRAN program based on the grid 
published by \citet{md85}. Written in 1985 by T.T.\ Moon of the University London and modified in 1992 and 1997 by R.\ Napiwotzki of 
Universitaet Kiel \citep*[see ][]{n93}.} and 21\,154~K using the calibration of \citet{SJ93}. The close agreement between these $T_{\rm 
eff}$ values is due to the fact that the three temperature calibrations rely heavily on the OAO-2 absolute flux calibration of 
\citet{cod}. Taking a straight mean of the above three values we arrive at $T_{\rm eff,1} =21\,200$~K. Realistic standard deviations of 
the effective temperatures of early-type stars, estimated from the uncertainty of the absolute flux calibration, amount to about 3\,\% 
\citep{n93,j94} or 640~K for the $T_{\rm eff,1}$ in question, so that $\log T_{\rm eff,1} ={}$4.326$\pm$0.013. The standard deviation 
of $BC_1$ we estimate to be 0.20\,mag.

The revised {\em Hipparcos\/} parallax of $\pi^5$ Ori is equal to 2.43$\,\pm\,$0.39~mas \citep*{vL}. Taking the star's $V$ magnitude 
from \citet{m91}, $E(B-V)$ from the first paragraph of this section, and assuming $R_V ={}$3.2, we get $M_V ={}-$4.54$^{+\rm 
0.32}_{-\rm 0.38}$~mag, $M_{\rm bol} ={}-$6.68$^{+\rm 0.38}_{-\rm 0.43}$~mag, and $\log L_1/{\rm L}_{\sun} ={}$4.57$^{+\rm 0.17}_{-\rm 
0.15}$. In computing $\log L_1/$L$_{\sun}$, we assumed M$_{{\rm bol}{\sun}} ={}$4.74~mag, a value consistent with BC$_{\sun} 
=-{}$0.07~mag and $V_{\sun} ={}-$26.76~mag \citep{torr}.

\begin{figure} 
\includegraphics[width=\columnwidth]{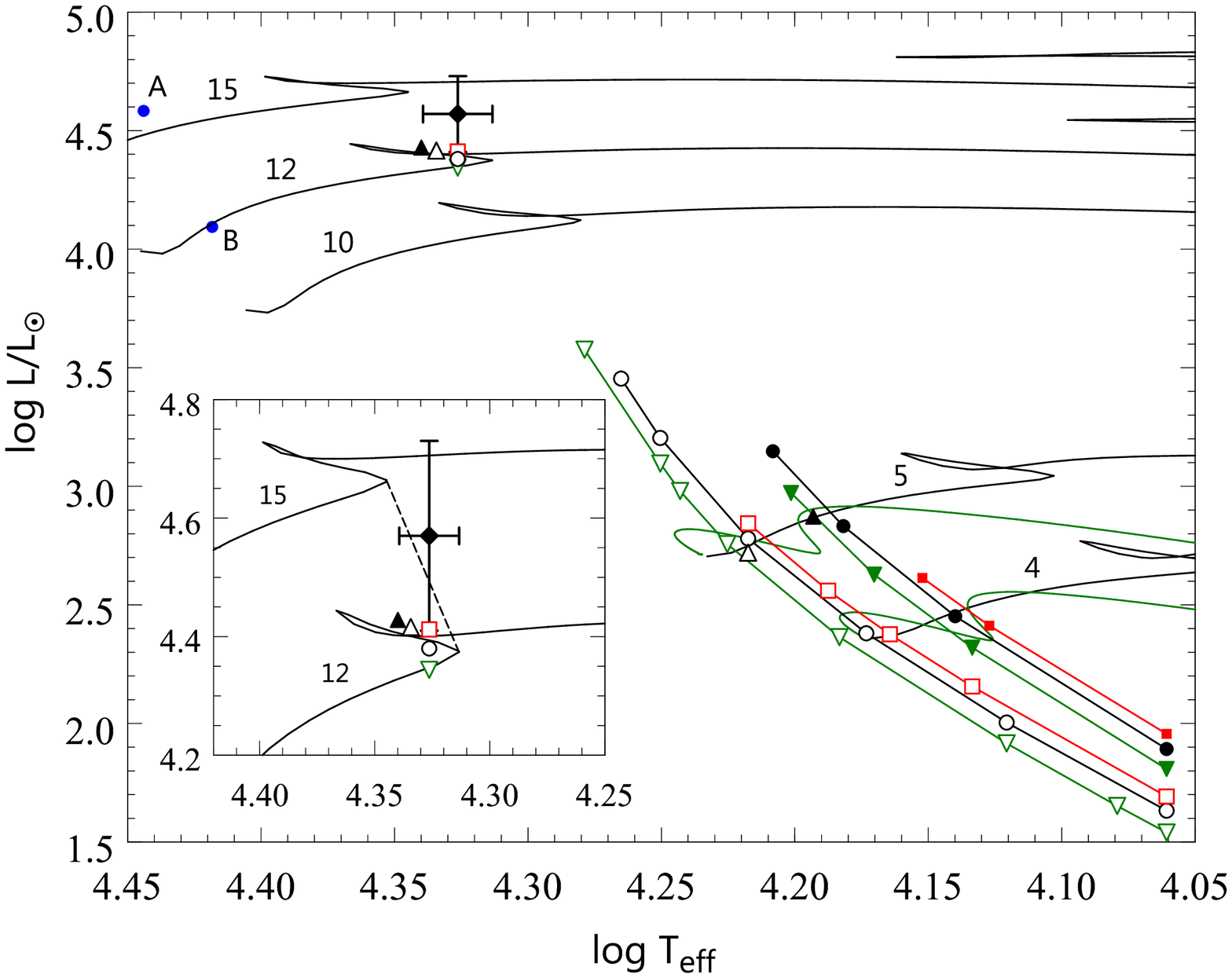} 
\caption{The components of $\pi^5$ Ori in the HR diagram. The diamond with error bars is plotted using $\log (T_{\rm eff,1} 
={}$21\,200~K$)$ and $\log L_1/{\rm L}_{\sun}$ derived in Section~\ref{parameters} under assumption of negligible brightness of the 
secondary component. The green open inverted-triangles, black open circles and red open squares are from the W-D solutions for $M_1 
={}$11, 12 and 13~M$_{\sun}$, respectively, $T_{\rm eff,1} ={}$21\,200~K, a radiative-envelope bolometric albedo $\alpha_1 = \alpha_2 
={}$1.0 and a range of $\log T_{\rm eff,2}$, while the green filled inverted-triangles, black filled circles and red filled squares are 
from the W-D solutions with the same $M_1$, $T_{\rm eff,1}$ and $\alpha_1$ as above, a convective-envelope bolometric albedo $\alpha_2 
={}$0.5 and a range of $\log T_{\rm eff,2}$. The $\log T_{\rm eff,2}$ ranges are specified in Section~\ref{WDsolutions} where the W-D 
solutions are discussed. The primary component's HR diagram positions from the W-D solutions (open inverted-triangle, open circle and open 
square at upper left and in the inset) were virtually unaffected by the assumed values of the secondary's albedo and effective 
temperature. The W-D duplicity-corrected positions of both components (black open and filled triangles for $\alpha_2 ={}$1.0 and 0.5, 
respectively) are plotted using the parameters from Table~\ref{TabB3}; the filled triangle representing the primary component (upper 
left and the inset) was shifted by 0.004 dex to the left to avoid overlap. The blue dots represent components of the detached eclipsing 
binary V453 Cyg plotted using the data from table 1 of \citet{TAG10} to be referred to in the last paragraph of 
Section~\ref{WDsolutions}. Also plotted are the 4, 5, 10, 12 and 15\,M$_{\sun}$ Padova evolutionary tracks for $Y ={}$0.26 and $Z 
={}0.017$ (black lines) from \citet{Ber+09}, and the 4 and 5\,M$_{\sun}$ Pisa pre-MS tracks from \citet{Tog+2011} for $Y ={}$0.265, $Z 
={}$0.0175 and the mixing-length parameter of 1.68\,$H_{\rm p}$, where $H_{\rm p}$ is the pressure scale height (dark-green lines). In 
the inset, the TAMS is indicated  (short-dashed line).} 
\label{FigA1} 
\end{figure}

In Fig.~\ref{FigA1}, $\pi^5$ Ori is plotted in the HR diagram together with the 4, 5, 10, 12 and 15\,M$_{\sun}$ Padova evolutionary 
tracks from \citet{Ber+09} for $Y ={}$0.26 and $Z=0.017$, and the 4 and 5\,M$_{\sun}$ Pisa pre-MS tracks from \citet*{Tog+2011} for $Y 
={}$0.265, $Z ={}$0.0175 and the mixing-length parameter of 1.68\,$H_{\rm p}$, where $H_{\rm p}$ is the pressure scale height. As can 
be seen from the figure (see the inset), the star falls to the right and above  the terminal main-sequence (TAMS) but is off the region 
corresponding to the late hydrogen-burning (HB) evolutionary stage by less than 1$\sigma$ in $\log T_{\rm eff}$ and in $\log L/{\rm 
L}_{\sun}$. The green inverted triangles, black circles and red squares (open and filled, connected with straight lines and otherwise) 
are from the W-D solutions discussed in Section~\ref{WDsolutions}.

\begin{figure} 
\includegraphics[width=\columnwidth]{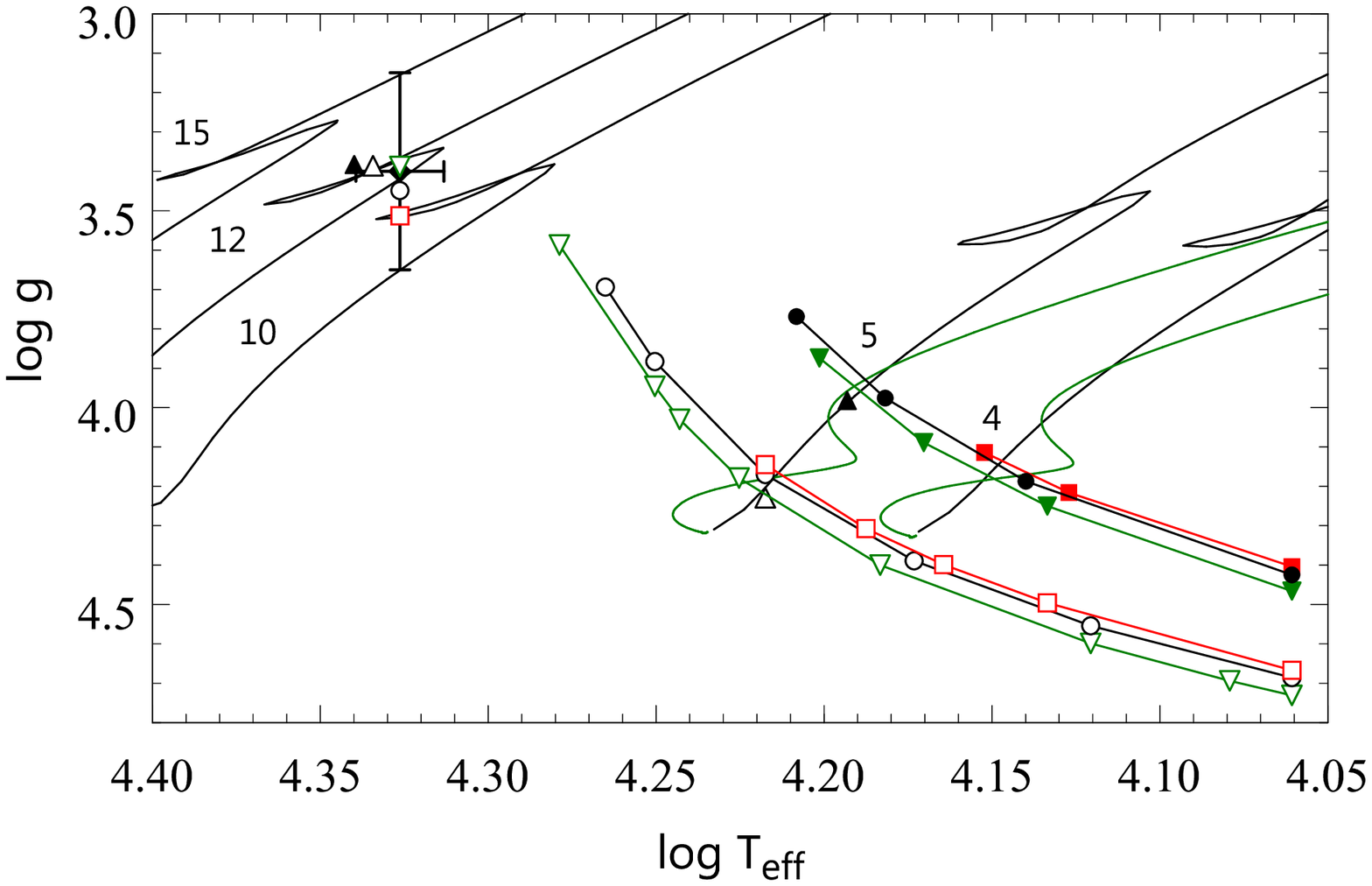} 
\caption{The components of $\pi^5$ Ori in the $\log T_{\rm eff}$, $\log g$ plane. The symbols are the same as in Fig.~\ref{FigA1} 
except that the black open circle and red open square at upper left, representing the position of the primary from the 12 and 
13~M$_{\sun}$ W-D solutions, were shifted downwards by 0.06 and 0.12~dex, respectively, to avoid overlap. Also plotted are the 
evolutionary tracks from the sources referenced in the caption to Fig.~\ref{FigA1}.}
\label{FigA2} 
\end{figure}

The surface gravity of a B-type star can be obtained from its $\beta$ index. There are two values of the $\beta$ index of $\pi^5$ Ori 
in the literature: 2.603 \citep{hm} and 2.597~mag \citep{p15}. From a straight mean of these numbers and $T_{\rm eff}$, we get $\log g 
={}$3.40 using the $T_{\rm eff}$, $\beta$ grid of \citet{SD95} modified by \citet{DzJ99}. According to \cite{n93}, the uncertainty of 
the $\beta$-index surface gravities of hot stars is equal to 0.25~dex; we shall adopt this value as the standard deviation of the 
star's $\log g$. Using the above derived $T_{\rm eff} ={}$21\,200$\,\pm\,$640~K and $\log g ={}$3.40$\,\pm\,$0.25, we plot $\pi^5$ Ori 
in Fig.~\ref{FigA2} together with the MS and pre-MS evolutionary tracks, and $\log g$ resulting from the W-D modeling to be discussed 
in Section~\ref{WDsolutions}.

The 2016 version of the {\sc PASTEL} catalogue \citep{Soub+16} lists $T_{\rm eff} ={}$21\,860~K and $\log g ={}$3.51 obtained by 
\cite{GL92} from Str\"omgren colour indices and H$\gamma$ line profiles through a comparison with colours and line profiles from 
Kurucz line-blanketed atmospheres. These values agree quite well with those we derived: the  former is greater than ours by slightly 
more than 1$\sigma$, while the latter, by less than 0.5$\sigma$. 

\section{The W-D Modeling}\label{WDsolutions}

\begin{figure} 
\includegraphics[width=\columnwidth]{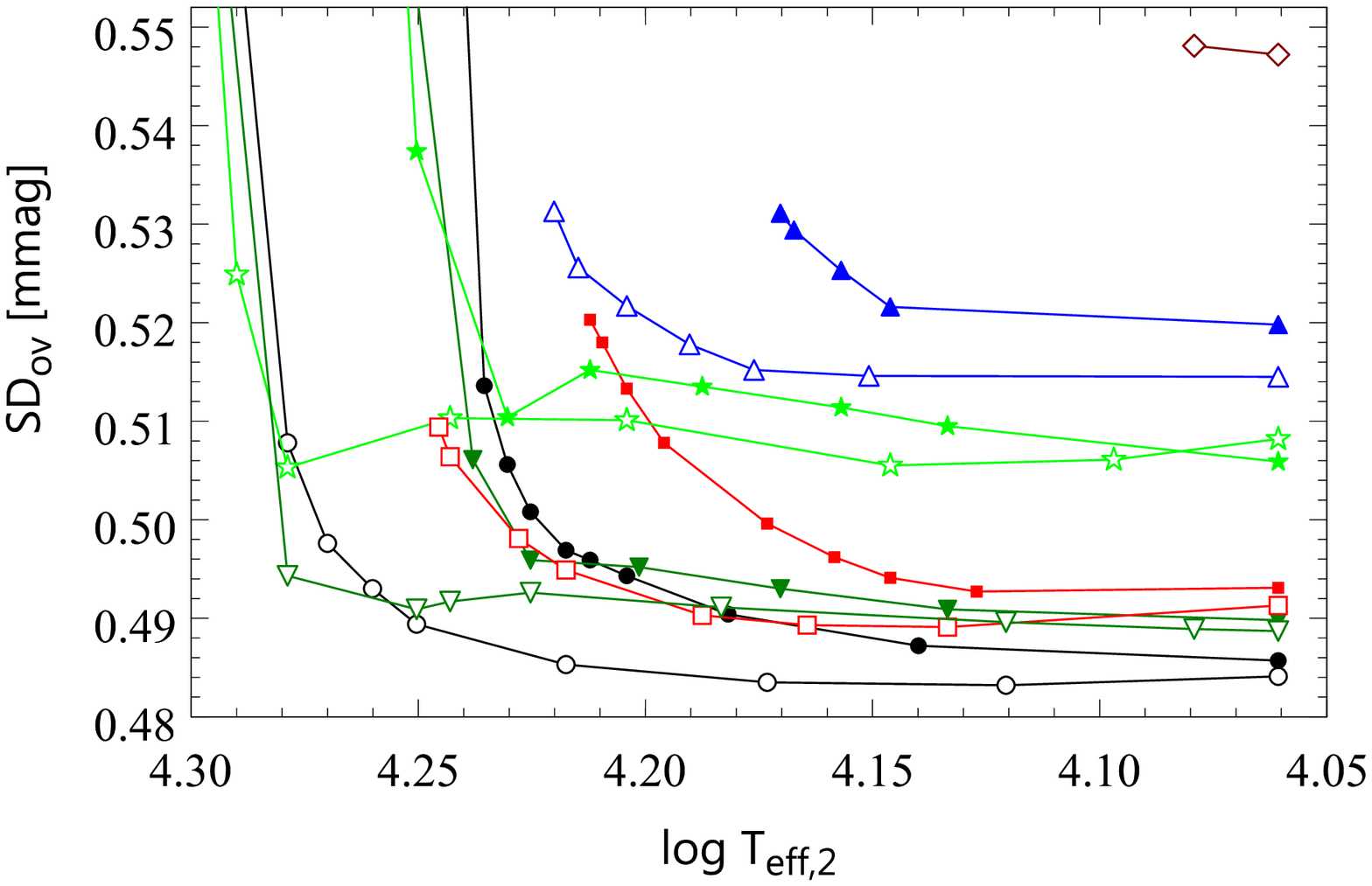} 
\caption{The overall standard deviation, $SD_{\rm ov}$, of the W-D fits to the observed light-curves of Fig.~\ref{Fig03} for $T_{\rm 
eff,1} ={}$21\,200~K and $M_1 ={}$10 (light-green asterisks), 11 (dark-green inverted triangles), 12 (black circles), 13 (red squares), 
14 (blue triangles) and 15~M$_{\sun}$ (brown diamonds). Empty symbols represent solutions with $\alpha_2 ={}$1.0, filled symbols, with 
$\alpha_2 ={}$0.5. For $M_1 ={}$13, 14 and 15~M$_{\sun}$, the primary component exceeded its critical lobe in the solutions on the 
left-hand side of the last plotted point; for $M_1 ={}$15~M$_{\sun}$ and $\alpha_2 ={}$0.5, the primary component exceeded its critical 
lobe if $\log T_{\rm eff,2} \geq{}$4.03.}
\label{FigB1} 
\end{figure}

The light-curves shown in Fig.~\ref{Fig03} as dots were subject to modeling by means of the 2015 version of the Wilson-Devinney code 
\citep[hereafter W-D, ][]{WiDe71,Wi79}. In the modeling, we used \citet{Mi50} orbital period of 3.700373~d and the semi-amplitude of 
the RV curve $K_1 ={}$58.4~km\,s$^{-1}$ obtained in Section~\ref{orbital-curves} from the combined observations of \citet*{Lee13} and 
\citet*{Mi50} assuming zero eccentricity. For both components, the limb darkening coefficients were taken from the logarithmic-law 
tables of Walter V.~Van Hamme\footnote{http://faculty.fiu.edu/{$\sim$}vanhamme/limb-darkening/ , see also \citet{VH}.}. We assumed 
[M/H] $= 0$ and used $\lambda$421 and 620.5\,nm monochromatic coefficients for the blue and red data, respectively.  In treating the 
reflection effect, we used the detailed model with six reflections (MREF $=2$, NREF $=6$). The reflection effect is small but 
significant: it accounts for the difference in the depth of the minima seen in Fig.~\ref{Fig03}. Under assumption of synchronous 
rotation, the observed $V_{\rm rot} \sin i ={}$90~km\,s$^{-1}$ \citep{GG05} and the radius of the primary component yield the 
inclination of the orbit. For a given $M_1$, one then gets $M_2$. Guided by the position of the star in the HR diagram in relation to 
the evolutionary tracks (see Fig.~\ref{FigA1}), we assumed $M_1 \leq{}$15~M$_{\sun}$ and then computed W-D solutions for $M_1 ={}$10, 
11, 12, 13, 14 and 15~M$_{\sun}$, a $\pm$640~K range of $T_{\rm eff,1}$ around $T_{\rm eff,1} ={}$21\,200~K, the value derived in 
Section~\ref{parameters}, a number of values of $T_{\rm eff,2}$, and the primary's radiative-envelope bolometric albedo $\alpha_1 
={}$1.0. Since the evolutionary state of the secondary is not known, we computed two series of solutions, one with the secondary's 
bolometric albedo $\alpha_2 ={}$1.0, and the other,  with a convective-envelope bolometric albedo $\alpha_2 ={}$0.5. We found that  the 
overall standard deviation, $SD_{\rm ov}$, of the W-D fit to the observed light-curves is a function of $M_1$ and $T_{\rm eff,2}$. This 
result is set out in Fig.~\ref{FigB1} with $\log T_{\rm eff,2}$ as the abscissa. As can be seen from the figure, the best fits are 
obtained for $M_1 ={}$12~M$_{\sun}$, $\alpha_2 ={}$1.0, $\log T_{\rm eff,2} \leq{}$4.22, and $\alpha_2 ={}$0.5, $\log T_{\rm eff,2} 
\approx{}$4.06. For $M_1 ={}$10, 14 and 15~M$_{\sun}$, the fits are much less satisfactory. The parameters' ranges from the solutions 
which yield fits with $SD_{\rm ov} \leq 0.495$~mmag are listed in Table B1. The parameters of these solutions were used to plot the 
components in Figs.~\ref{FigA1} and \ref{FigA2}. As can be seen from Table~\ref{TabB1}, the primary's W-D radius and luminosity are not 
sensitive to the secondary's effective temperature and albedo, so that for given $M_1$ and $T_{\rm eff,1}$, $R_1$, $M_2$ and the 
primary's HR diagram position remain nearly unchanged. In contrast, $R_2$ and the HR diagram position of the secondary vary strongly 
with $T_{\rm eff,2}$. We shall take advantage of the last property in the next paragraph. 

\begin{table*}
\caption{The parameters of the W-D solutions which yield fits to the observed light-curves of Fig.~\ref{Fig03} with the overall 
standard deviation $SD_{\rm ov} \leq 0.495$~mmag.}
\label{TabB1} 
\begin{center}
  \begin{tabular}{@{}crrcrrcccc@{}}
  \hline
    \multicolumn{1}{c}{$M_1$} &\multicolumn{1}{c}{$\alpha_2$}& \multicolumn{1}{c}{$R_1$} & 
\multicolumn{1}{c}{$i$}&\multicolumn{1}{c}{$\log L_1/$L$_{\sun}$} & \multicolumn{1}{c}{$\log g_1$} & 
\multicolumn{1}{c}{$\log T_{\rm eff,2}$}& \multicolumn{1}{c}{$R_2$} &\multicolumn{1}{c}{$\log L_2/$L$_{\sun}$} & 
\multicolumn{1}{c}{$\log g_2$}  \\ 
\multicolumn{1}{c}{[M$_{\sun}$]} & & \multicolumn{1}{c}{[R$_{\sun}$]} & & & & & \multicolumn{1}{c}{[R$_{\sun}$]} &  & \\
  \hline
11&1.0&11.0-11.4&36\fdg6-35\fdg3&4.34-4.37&3.39-3.36&4.061-4.279&1.50-5.70&1.54-3.58&4.73-3.59\\
11&0.5&11.1-11.3&36\fdg5-35\fdg8&4.34-4.36&3.39-3.38&4.061-4.201&2.04-4.06&1.81-2.97&4.47-3.87\\
12&1.0&11.5-11.8&34\fdg7-33\fdg8&4.38-4.40&3.39-3.37&4.061-4.265&1.66-5.30&1.63-3.45&4.69-3.69\\
12&0.5&11.5-11.7&34\fdg8-34\fdg2&4.38-4.39&3.39-3.38&4.061-4.208&2.24-4.82&1.89-3.15&4.42-3.77\\
13&1.0&12.0-12.1&33\fdg3-33\fdg0&4.41-4.42&3.40-3.39&4.061-4.218&1.78-3.26&1.69-2.84&4.67-4.14\\
13&0.5&12.0-12.1&33\fdg3-33\fdg1&4.41-4.42&3.39-3.39&4.061-4.152&2.41-3.38&1.96-2.61&4.40-4.11\\
  \hline
\end{tabular}
\end{center}
\end{table*}

Since the magnitude difference between the components of $\pi^5$~Ori is not known, we cannot correct the parameters of the primary 
component derived in Section~\ref{parameters} from the combined-light magnitude and colour indices for the light dilution caused by the 
secondary. However, using magnitude differences provided by the W-D solutions we can compute duplicity corrections for a given $T_{\rm 
eff,1}$ (uncorrected), $M_1$ and $T_{\rm eff,2}$. As an example, we chose the $T_{\rm eff,1} ={}$21\,200~K, $M_1 ={}$12~M$_{\sun}$, 
$\alpha_2 ={}$1.0 and 0.5 solutions with $T_{\rm eff,2}$ selected in such a way that the evolutionary masses estimated from the 
evolutionary tracks shown in Fig.~\ref{FigA1} were equal to $M_2$, viz.\ $T_{\rm eff,2} ={}$16\,400~K for $\alpha_2 ={}$1.0 and the MS 
Padova tracks, and $T_{\rm eff,2} ={}$15\,400~K for $\alpha_2 ={}$0.5 and the pre-MS Pisa tracks. Taking the blue ($\lambda$421\,nm) 
and red ($\lambda$620.5\,nm) magnitude differences from these solutions we obtained the $V$ ($\lambda$555\,nm) magnitude difference 
$\Delta V ={}$3.2 and 2.8~mag for $\alpha_2 ={}$1.0 and 0.5, respectively. Assuming luminosity class V for the secondary, we estimated 
its spectral type from the tables of \cite{Lang} to be B6.7 and B5.3 for $\Delta V ={}$3.2 and 2.8~mag, respectively. Then, from the 
average values of $c_0$ and $m_0$ as a function of MK type and the average values of $(b-y)_0$ as a function of $c_0$ \citep*[tables II 
and I of][]{c78} we obtained the duplicity corrections (to be subtracted from the combined $c_0$) of 0.015 and 0.018~mag for $\Delta V 
={}$3.2 and 2.8~mag, respectively. In terms of $T_{\rm eff}$, the correction (to be added to the observed value) is 390 and 470~K, 
respectively. The duplicity correction to $\beta$ was computed assuming that for single stars $\beta_{\rm wide}$ scales as the 
magnitude at 486~nm. Assuming again luminosity class V for the secondary, we then get the duplicity correction (to be subtracted from 
the combined $\beta$\/) of 0.005 and 0.007~mag for $\Delta V ={}$3.2 and 2.8~mag. Consequently, the corrections to be subtracted from 
$\log g$ obtained from the combined $\beta$ and $c_0$ are equal to 0.07 and 0.03~dex for $\Delta V ={}$3.2 and 2.8~mag, 
respectively. The corrections for the different $\Delta V$ differ because the duplicity-corrected $c_0$ differ. Finally, the 
corrections to $\log L/$L$_{\sun}$ (to be subtracted from the uncorrected values), with the corrections to $BC$ taken into account, 
were equal to 0.006 and 0.014~dex for $\Delta V ={}$3.2 and 2.8~mag, respectively. The duplicity-corrected photometric indices and 
fundamental parameters of the primary component are listed in Table~\ref{TabB2}, and its duplicity-corrected positions are shown in 
Figs.~\ref{FigA1} and \ref{FigA2} as the triangles at upper left. With the duplicity-corrected $T_{\rm eff,1}$, we obtained solutions 
for $M_1 ={}$12~M$_{\sun}$ for which $M_2$ were equal to the evolutionary masses. The parameters of these solutions are listed in 
Table~\ref{TabB3}. Note that the primary's W-D luminosities are lower than the duplicity-corrected value by slightly less than 
1$\sigma$. 

The problem with the above example is that the evolutionary ages do not match: the TAMS age on the 12~M$_{\sun}$ track is equal to 
18~Myr while the evolutionary ages on the 5~M$_{\sun}$ tracks are equal to 25~Myr for the secondary component on the MS track 
($\alpha_2 ={}$1.0), and 0.8~Myr for the secondary on the pre-MS track ($\alpha_2 ={}$0.5). The 18~Myr evolutionary age of the 
$\alpha_2 ={}$1.0 secondary would result if we shifted the 5~M$_{\sun}$ MS track by $-$0.013 dex in $\log T_{\rm eff}$ and by $-$0.11 
dex in $\log L/$L$_{\sun}$. A similar result would be obtained by appropriately shifting the HR diagram position of the secondary. In 
view of the uncertainties of our data (e.g.~those of the evolutionary tracks on the theoretical side, and $V_{\rm rot} \sin i$ on the 
observational side) the mismatch of the components' evolutionary masses for the $\alpha_2 ={}$1.0 solution is tolerable. However, it is 
certainly not for the $\alpha_2 ={}$0.5 solution. Thus, our example suggests that the secondary component is in the early stages of its 
MS evolution (open triangle at lower right in Figs.~\ref{FigA1} and \ref{FigA2}). This conclusion is in keeping with the fact, seen in 
Fig.~\ref{FigB1}, that the $SD_{\rm ov}$ for the $\alpha_2 ={}$1.0 solutions are lower than those for the $\alpha_2 ={}$0.5 solutions. 

\begin{table}
\caption{The duplicity-corrected photometric indices and fundamental parameters of the primary component of $\pi^5$ Ori for the two 
values of the magnitude difference between the components, $\Delta V$, obtained from the $M_1 ={}$12~M$_{\sun}$, $T_{\rm eff,1} 
={}$21\,200~K W-D solutions discussed in Section~\ref{WDsolutions}.}
\label{TabB2} 
\begin{center}
 \begin{tabular}{@{}rrrrccc@{}}
  \hline
    \multicolumn{1}{c}{$\alpha_2$}&\multicolumn{1}{c}{$\Delta V$}&\multicolumn{1}{c}{$c_0$}&\multicolumn{1}{c}{$\beta$} & 
    \multicolumn{1}{c}{$T_{\rm eff,1}$}& \multicolumn{1}{c}{$\log L_1/$L$_{\sun}$} & \multicolumn{1}{c}{$\log g_1$} 
\\
  \hline
1.0&3.2&0.110&2.595&21\,590$\pm$650&4.564$\pm$0.16&3.33$\pm$0.25\\
0.5&2.8&0.107&2.593&21\,670$\pm$650&4.556$\pm$0.16&3.36$\pm$0.25\\
  \hline
\end{tabular}
\end{center}
\end{table}

\begin{table*}
\caption{The parameters of the components of $\pi^5$ Ori obtained from the W-D solutions for $M_1 ={}$12~M$_{\sun}$ and the 
duplicity-corrected $T_{\rm eff,1}$ listed in Table~\ref{TabB2}.}
\label{TabB3} 
\begin{center}
  \begin{tabular}{@{}rccccccccccc@{}}
  \hline
    \multicolumn{1}{c}{$\alpha_2$}& \multicolumn{1}{c}{$R_1$} & \multicolumn{1}{c}{$i$} &\multicolumn{1}{c}{$a$} 
& \multicolumn{1}{c}{$M_2$}& \multicolumn{1}{c}{$\log T_{\rm eff,2}$}&\multicolumn{1}{c}{$R_2$}&
\multicolumn{1}{c}{$\log L_1/$L$_{\sun}$} & \multicolumn{1}{c}{$\log L_2/$L$_{\sun}$}& \multicolumn{1}{c}{$\log g_1$} & 
\multicolumn{1}{c}{$\log g_2$} & \multicolumn{1}{c}{$SD_{\rm ov}$} \\
    & \multicolumn{1}{c}{[R$_{\sun}$]} &  &\multicolumn{1}{c}{[R$_{\sun}$]} 
& \multicolumn{1}{c}{[M$_{\sun}$]}& &\multicolumn{1}{c}{[R$_{\sun}$]} & & & & 
 & \multicolumn{1}{c}{[mmag]} \\
  \hline
1.0&11.6&34\fdg4&25.9&4.95&4.218&2.83&4.42&2.72&3.39&4.24&0.484\\
0.5&11.7&34\fdg3&25.9&4.96&4.193&3.76&4.43&2.87&3.38&3.98&0.488\\
  \hline
\end{tabular}
\end{center}
\end{table*}

\begin{figure} 
\includegraphics[width=\columnwidth]{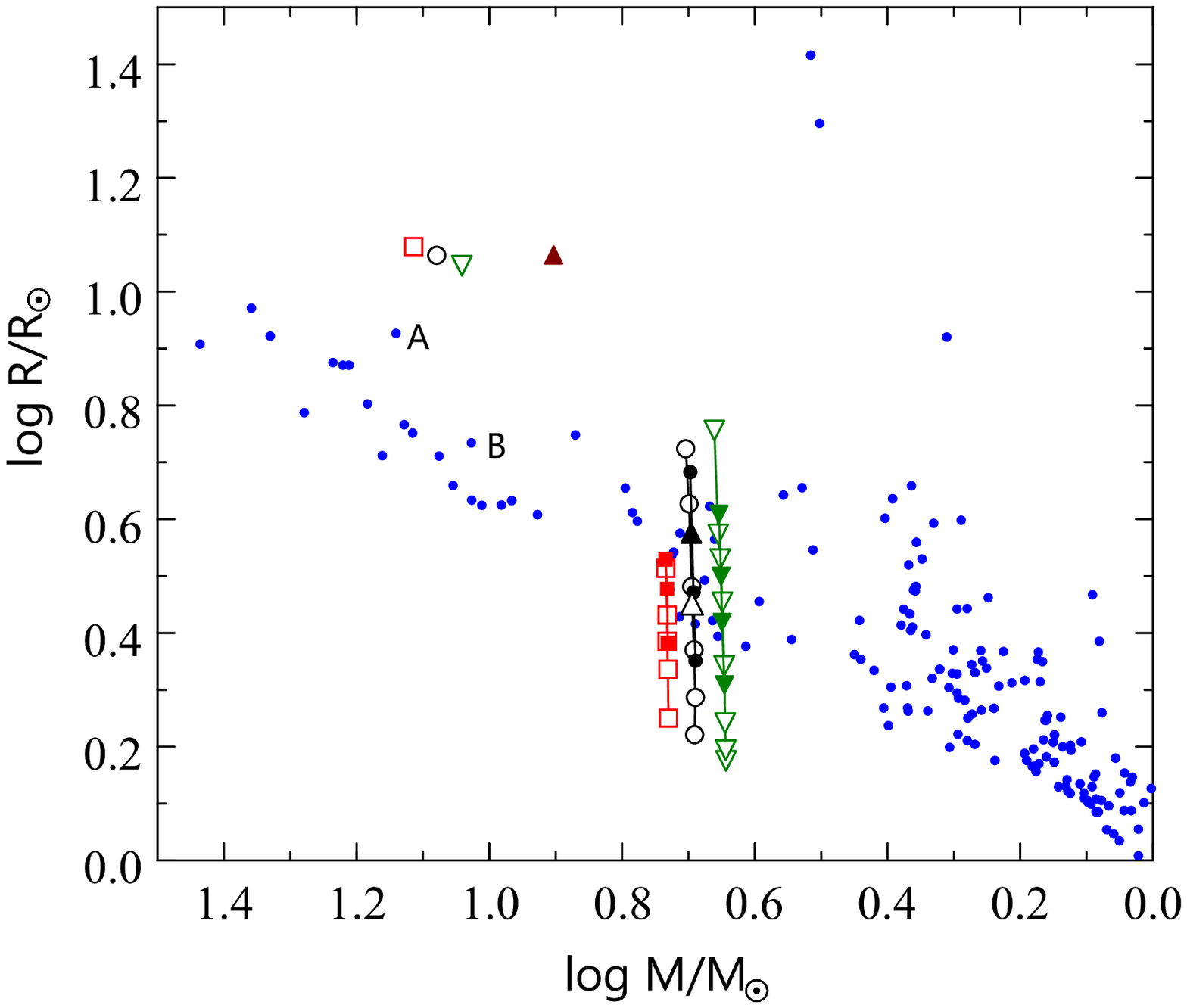} 
\caption{The radii of the components of $\pi^5$ Ori, obtained from the W-D solutions with $T_{\rm eff,1} ={}$21\,200~K, $M_1 ={}$11, 12 
and  13~M$_{\sun}$ (green inverted triangles, black circles and red squares, respectively), the same as those used in plotting 
Figs.~\ref{FigA1} and \ref{FigA2}, compared with the empirical masses and radii of the SB2 eclipsing binaries from table 1 of 
\citet{TAG10} (blue dots; those labeled A and B represent components of the detached eclipsing binary V453 Cyg). The primary 
component's radii (open inverted triangle, open circle and open square at upper left) were unaffected by the assumed value of the 
secondary's bolometric albedo and effective temperature. The secondary's radii from the $M_1 ={}$12~M$_{\sun}$ and duplicity-corrected 
$T_{\rm eff,1}$ solutions, listed in the seventh column of Table~\ref{TabB3}, are shown as black open and filled triangles for 
$\alpha_2 ={}$1.0 and 0.5, respectively; the primary's radii from these solutions are not plotted because they would coincide with the 
open circle at upper left. The brown triangle is from \citet{Mo85} solution with negligible brightness of the secondary component; 
$\log R_1/$R$_{\sun}$ from his other solution differs from that shown by an insignificant 0.05.} 
\label{FigB2} 
\end{figure}

The radii of the components of $\pi^5$ Ori derived from the $M_1 ={}$11, 12 and 13~M$_{\sun}$ W-D solutions given in Table~\ref{TabB1} 
are compared in Fig.~\ref{FigB2} with the empirical masses and radii of the SB2 eclipsing binaries from table 1 of \citet{TAG10}. As 
can be seen from the figure, good agreement of the W-D radii of the secondary component with the empirical ones is obtained over the 
whole interval of $\log T_{\rm eff,2}$ listed in Table~\ref{TabB1} for $\alpha_2 ={}$0.5 (filled inverted triangles, filled circles and 
filled squares), and over the  $\log T_{\rm eff,2}$ intervals [4.17,4.27], [4.16,4.26] and [4.14,4.22] for $\alpha_2 ={}$1.0, $M_1 
={}$11, 12 and 13~M$_{\sun}$, respectively (open inverted triangles, open circles and open squares). The secondary's radii from the 
$M_1 ={}$12~M$_{\sun}$ and duplicity-corrected $T_{\rm eff,1}$ solutions given in Table~\ref{TabB3} (black open and filled triangles) 
fall within the [4.16,4.26] interval. However, the primary's W-D radii are much greater than the empirical ones of similar mass. In 
particular, they are greater than the greatest empirical radius in the 10 to 20~M$_{\sun}$ mass range, viz.\ that of the primary 
component of V453 Cyg. The explanation is trivial: as can be seen from Fig.~\ref{FigA1}, the primary component of $\pi^5$ Ori is in a 
more advanced stage of evolution than V453 Cyg. It can be easily verified using the data from table 1 of \citet{TAG10} that the 
components of the remaining SB2 eclipsing binaries in the same mass range are even younger. Explaining the large \citet{Mo85} $R_1$ 
(brown triangle) in a similar fashion is problematic because an 8~M$_{\sun}$ primary of that radius would be well into the shell 
hydrogen-burning evolutionary stage. \citet{Mo85} solutions are unfeasible for yet another reason: the 8~M$_{\sun}$ W-D light-curves 
fit those observed with $SD_{\rm ov} >{}$0.57~mmag, a value greater than those plotted in Fig.~\ref{FigB1}. 

\bsp
\label{lastpage}
\end{document}